\documentclass[aps,amsmath,amssymb, notitlepage,
twocolumn,
nofootinbib,
superscriptaddress
]{revtex4-1}

\usepackage{amsmath}
\usepackage{amssymb}
\usepackage{amsthm}
\usepackage{amsfonts}

\usepackage[all]{xy}
\usepackage{graphicx}
\usepackage{braket}

\usepackage{color}
\definecolor{darkblue}{rgb}{0.,0.,0.4}
\definecolor{darkred}{rgb}{0.5,0.,0.}
\usepackage[pdftex,colorlinks=true,linkcolor=darkblue,citecolor=blue,urlcolor=darkred]{hyperref}

\newcommand{\beq}{\begin{equation}}
\newcommand{\eeq}{\end{equation}}
\newcommand{\mc}[1]{\mathcal{#1}}
\newcommand{\la}{\langle}
\newcommand{\ra}{\rangle}
\newcommand{\half}{\frac{1}{2}}

\begin{document}

\title{Bulk characterization of topological crystalline insulators: stability under interactions and relations to symmetry enriched U(1) quantum spin liquids}

\author{Liujun Zou}
\affiliation{Department of Physics, Harvard University, Cambridge, Massachusetts, 02138, USA}
\affiliation{Department of Physics, Massachusetts Institute of Technology, Cambridge, Massachusetts, 02138, USA}

\begin{abstract}
Topological crystalline insulators (TCIs) are nontrivial quantum phases of matter protected by crystalline (and other) symmetries. They are originally predicted by band theories, so an important question is their stability under interactions. In this paper, by directly studying the physical bulk properties of several band-theory-based nontrivial TCIs that are conceptually interesting and/or experimentally feasible, we show they are stable under interactions. These TCIs include: (1) a weak topological insulator, (2) a TCI with a mirror symmetry and its time reversal symmetric generalizations, (3) a doubled topological insulator with a mirror symmetry, and (4) two TCIs with symmetry-enforced-gapless surfaces. We describe two complementary methods that allow us to determine the properties of the magnetic monopoles obtained by coupling these TCIs to a $U(1)$ gauge field. These methods involve studying different types of surface states of these TCIs. Applying these methods to our examples, we find all of them have nontrivial monopoles, which proves their stability under interactions. Furthermore, we discuss two levels of relations between these TCIs and symmetry enriched $U(1)$ quantum spin liquids (QSLs). First, these TCIs are directly related to $U(1)$ QSLs with crystalline symmetries. Second, there is an interesting correspondence between $U(1)$ QSLs with crystalline symmetries and $U(1)$ QSLs with internal symmetries. In particular, the TCIs with symmetry-enforced-gapless surfaces are related to the ``fractional topological paramagnets" introduced in Ref. \onlinecite{Zou2017} by Zou {\it et al}.
\end{abstract}

\date{\today}

\maketitle

\tableofcontents

\section{Introduction} \label{sec: introduction}

\begin{figure}
  \centering
  \includegraphics[width=0.5\textwidth]{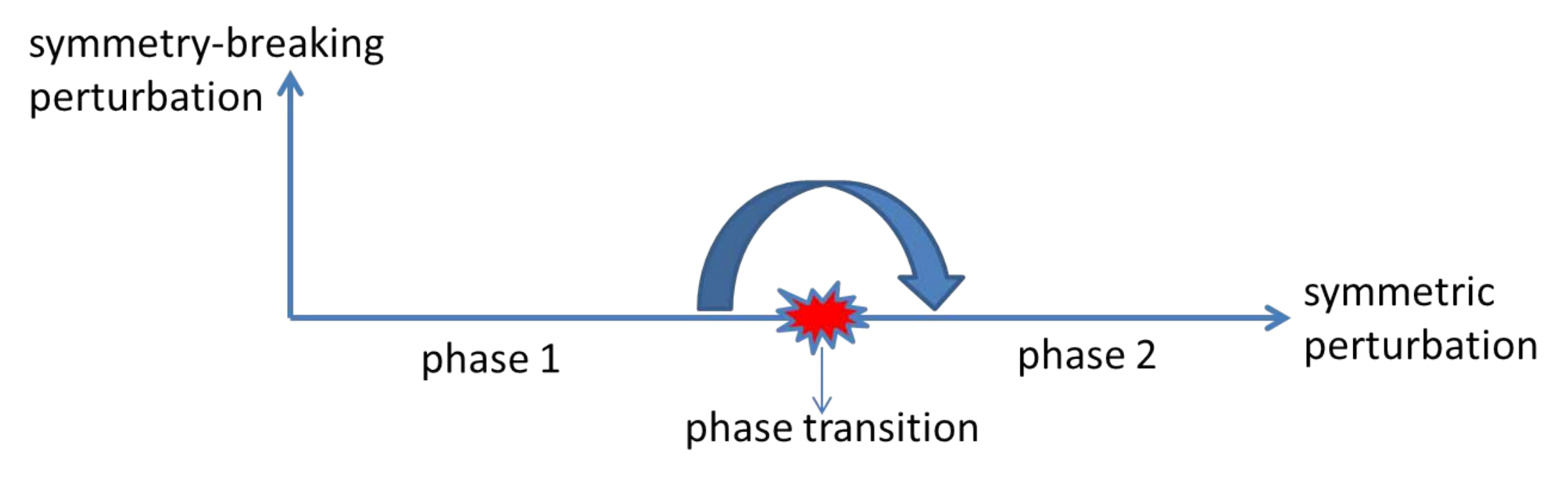}\\
  \caption{The notion of symmetry protected distinction of quantum phases of matter. As long as the relevant symmetries are preserved, the two phases cannot be connected without crossing a phase transition. However, if symmetry-breaking perturbations are allowed, the phase transition can be avoided.}\label{fig: symmetry-protected-disctinction}
\end{figure}

The notion of symmetry protected distinction of quantum phases of matter is by now well appreciated: some quantum phases are smoothly connected to each other in the absence of symmetry, but when the relevant symmetries are present, these phases are sharply distinguished, {\it i.e.} it is not possible to go from one phase into the other without crossing a phase transition (see Figure \ref{fig: symmetry-protected-disctinction}). The most well-known example may be the distinction between topological insulators and trivial insulators: in the presence of charge conservation and time reversal symmetries, these two types of insulators are separated by a phase transition. However, once these symmetries are allowed to be broken, they can be smoothly connected.\cite{Hasan2010,Qi2010,Hasan2010a}

Topological crystalline insulators (TCIs) are another class of topological materials that have been under intense recent studies. These are short-range entangled phases that are protected by crystalline symmetries: they cannot (can) be connected to a trivial insulator in the presence (absence) of the relevant crystalline (and other) symmetries.\cite{Fu2010,Ando2015,Hughes2010, Po2017,Watanabe2017} These states are not only of conceptual interests, but the state-of-the-arts in synthesizing materials also offer great opportunities to realizing them experimentally.\cite{Hsieh2012,Tanaka2012,Dziawa2012,Xu2012, Rasche2013,Okada2013,Pauly2015, Wang2016,Alexandradinata2016,Ezawa2016,Ma2016}

The majority of the theoretical studies of TCIs are based on band theories, which are valid for systems made of non-interacting or weakly-interacting fermions. An important question then is whether the nontrivial TCIs predicted by band theories are stable under interactions, or, more precisely, whether these band-theory-based nontrivial TCIs can be smoothly connected to a trivial insulator once strong interactions are switched on (still in the presence of the relevant symmetries). Very recently, the stability of various such TCIs under interactions has been considered and some of them are shown to be unstable, by studying the anomalous surface properties of the TCIs, studying certain dimensionally reduced versions of the TCIs, or studying the formal field theories of the TCIs.\cite{You2014,Isobe2015,Chiu2015,Morimoto2015,Shapourian2016, Song2016,Shiozaki2016, Song2016a,Thorngren2016,Huang2017, Cheng2017,Hong2017} However, an important and interesting question has remained unanswered: how to characterize nontrivial interacting TCIs directly by their bulk properties in a physical way?

In this paper we consider interacting three dimensional fermionic TCIs protected by crystalline symmetries together with a $U(1)$ charge conservation symmetry, and in some cases other symmetries (such as time reversal or $SU(2)$) are also present. Because all these insulators have a $U(1)$ symmetry, it is interesting and helpful to ask what happens if such TCIs are coupled to a (compact) dynamical $U(1)$ gauge field, and in particular, whether the resulting magnetic monopoles in the TCIs will be nontrivial. As a theoretical tool, this idea has provided great insights in other subjects where a $U(1)$ symmetry plays an important role,\cite{Qi2008, Wang2013,Wang2013a,Wang2013b,Wang2014,Wang2015,Wang2015a,Wang2015b, Metlitski2013,Metlitski2013a,Metlitski2015, Zou2017} although a magnetic monopole of the real electromagnetism has not been detected.

Therefore, we will discuss a particular type of bulk characterization of these TCIs: the properties of their magnetic monopoles once these TCIs are coupled to a dynamical $U(1)$ gauge field. We will focus on the cases where the crystalline symmetry responsible for the symmetry protection is either a translation symmetry or a mirror symmetry, and we will show several examples of band-theory-based nontrivial TCIs that have monopoles with nontrivial quantum numbers. Because these nontrivial quantum numbers of monopoles are rigid characterizations of the nontrivial TCIs, this implies that these TCIs will remain nontrivial even in the presence of strong interactions.

We will employ two complementary methods to derive the quantum numbers of monopoles, and both methods involve studying symmetric surface states of the TCIs. In general, there can be two types of symmetric surface states of a TCI: it can either be gapless, or it can be gapped at the cost of developing a surface topological order (STO). The two methods that we will be using are based on studying these two types of surface states. In the first method, we start from a symmetric gapless surface of the TCIs that can be described by free Dirac cones. Because the free Dirac theory is a conformal field theory (CFT), we can utilize state-operator-correspondence of a CFT to read off the quantum numbers the monopoles. In the second method, we start from a symmetric STO of the TCIs, then among the surface anyons of the STO we will identify an avatar of the bulk monopole, from which we can also extract the quantum numbers of the monopoles. Notice some TCIs may not have any symmetric STO, and we will discuss two such examples. That consistent results can be derived from different types of surface states is on the one hand assuring, and on the other hand indicating that it is usually a more direct and unified way to characterize a nontrivial TCI by its bulk physical properties.

The specific TCIs that we discuss are conceptually interesting and/or experimentally feasible, which include:

\begin{itemize}

\item[1.] Weak topological insulator (WTI).

A WTI is protected by charge conservation, time reversal and translation symmetries, and it can be viewed as many decoupled layers of two dimensional topological insulators.\cite{Hasan2010,Qi2010,Hasan2010a}

\item[2.] TCI protected by a mirror symmetry (MTCI).

This TCI has the same symmetries as the experimentally discovered TCI SnTe, and we will consider the ones that can have a symmetric surface with two Dirac cones. We also discuss two time reversal symmetric generalizations of this state (TMTCI-1 and TMTCI-2) in Appendix \ref{app: generalized-mirror-TCI}.

\item[3.] Doubled topological insulators protected by a mirror symmetry (DTI).

This example consists of two copies of three dimensional topological insulators further equipped with a mirror symmetry. It is well known that a single topological insulator is protected by charge conservation and time reversal, but two copies of them are smoothly connected to a trivial insulator if there is no other symmetry.\cite{Hasan2010,Qi2010,Hasan2010a} However, with a further mirror symmetry, the DTI can still be nontrivial.\cite{Watanabe2017}

\item[4.] TCIs with symmetry-enforced-gapless surfaces.

As the last examples, we construct two TCIs (SEG-1 and SEG-2) whose surfaces exhibit symmetry enforced gaplessness: their surfaces have to be gapless as long as the relevant symmetries are preserved, and a symmetric STO is disallowed even with strong interactions.\cite{Wang2014}

\end{itemize}

The studies of these TCIs are related to another intriguing subject. Recently there has been great interest in symmetry enriched long-range entangled phases: phases which cannot be smoothly connected to a trivial state even in the absence of any symmetry, and which acquire symmetry protected distinctions among themselves. In the last few years tremendous and systematic progress has been made in understanding two dimensional symmetry enriched long-range entangled states, \cite{Levin2012,NeupertSantosRyuEtAl2011,SantosNeupertRyuEtAl2011, EssinHermele2012,MesarosRan2012,HungWan2013,LuVishwanath2013,Wang2013, Barkeshli2014,Lan2016,Tarantino2015,Qi2017} and the studies of three dimensional symmetry enriched long-range entangled states just began.\cite{Xu2013a,ChenHermele2016,Ning2016} Very recently, in Ref. \onlinecite{Zou2017} the author (together with Chong Wang and T. Senthil) discussed, as one of the first systematic studies of three dimensional symmetry enriched long-range entangled phases, the classification and characterization of three dimensional symmetry enriched $U(1)$ quantum spin liquids (QSLs), which are long-range entangled phases with an emergent gapless photon at low energies (see also Ref. \onlinecite{Wang2015}). When the TCIs discussed in this paper is coupled to a (compact) dynamical $U(1)$ gauge field, the resulting state is precisely a $U(1)$ QSL enriched by a crystalline symmetry (together with some other symmetries in many cases). We discuss the properties of these crystalline symmetric $U(1)$ QSLs. Moreover, we demonstrate an interesting and potentially useful correspondence between crystalline-symmetry-enriched $U(1)$ QSLs discussed here and some internal-symmetry-enriched $U(1)$ QSLs discussed in Ref. \onlinecite{Zou2017}. In particular, the two TCIs with symmetry-enforced-gapless surfaces are found to be the crystalline-symmetric realizations of ``fractional topological paramagnets" introduced in Ref. \onlinecite{Zou2017}.

The rest of the paper is organized as follows. We first describe the two methods to derive the properties of monopoles in Sec. \ref{sec: monopole-property}. Then in Sec. \ref{sec: examples} we apply these methods to study the monopoles of the TCIs mentioned above, and we find that they all have nontrivial monopoles, so they are all stable under interactions. Next in Sec. \ref{sec: relation to SEU(1)QSL} we discuss the relations between the TCIs discussed in this paper and symmetry enriched $U(1)$ QSLs. Finally, we conclude with some discussions in Sec. \ref{sec: discussion}. The appendices contain some supplementary information.

\section{Properties of bulk monopoles from a symmetric surface state} \label{sec: monopole-property}

In this section we describe two complementary methods that allow us to read off the properties of monopoles of TCIs, and both methods utilize bulk-boundary correspondence. These methods can in principle be generalized and applied to study the properties of monopoles in other problems.

\subsection{A symmetric gapless surface with Dirac cones} \label{subsec: monopole-Dirac-surface}

We start with the case where a symmetric gapless surface of a TCI is known. This is usually the case if the TCI under interest can be realized by free fermions, where the symmetric gapless surface can be described by free Dirac cones. From the perspective of the free Dirac fermions on the surface, a monopole in the bulk corresponds to an instanton. Because free Dirac fermions are described by CFTs, the quantum numbers of their instantons can be read off by state-operator-correspondence.\cite{Borokhov2002} More precisely, the instanton operators correspond to the states obtained by threading a $2\pi$ flux through the surface theory. Guaranteed by the index theorem, each Dirac fermion will contribute a zero mode in this flux background. By studying how such states transform under certain symmetries, we can know how the instanton operators transform, and thus extract the quantum numbers of the monopoles. This procedure will be carried out for the examples, and here we just make some remarks.

First of all, we note for fermionic insulators, the monopoles must be bosonic. \cite{Wang2013b,Kravec2015} Therefore, we only need to pay attention to the possible nontrivial (projective) quantum numbers of the monopole under the relevant symmetries. This is not the case for bosonic TCIs, where the monopoles can in principle be fermionic, and one has to pay attention to both the statistics and quantum numbers of the monopole. In that case, as long as either the statistics or the quantum number is nontrivial, the monopole is nontrivial.

Second, if the symmetric surface of a band-theory-based nontrivial TCI is characterized by an odd number of Dirac cones, it is known that the monopole will carry half integer electric charge under the $U(1)$ symmetry (we will always measure the charge in units of the charge of electrons).\cite{Qi2008} This is always a nontrivial projective quantum number, so such TCIs will always be stable under interactions.

Therefore in this paper we focus on TCIs whose surfaces host two Dirac cones, where the monopole carries integer charge, which can then always be neutralized by attaching certain charge-1 fermions to it. If this neutral monopole carries nontrivial projective quantum numbers, this TCI is necessarily stable under interactions. Notice the inverse statement is not true. For example, some nontrivial band-theory-based TCIs with four surface Dirac cones have trivial monopoles, but they are still stable under interactions. These TCIs have bulk characterizations different than nontrivial monopoles, and they are beyond the scope of this paper.

\subsection{A symmetric surface topological order} \label{subsec: monopole-STO}

Now we describe how to derive the properties of the monopoles from a symmetric STO.\cite{Wang2013,Wang2013a,Wang2013b,Wang2014,Wang2015,Wang2015a, Wang2015b, Zou2017, Metlitski2013,Metlitski2013a,Metlitski2015} A topological order is characterized by a set of anyon types and their topological properties, such as fusion and braiding. If this topological order respects a $U(1)$ symmetry, the anyons will have definite $U(1)$ charges. Denote the $U(1)$ charge of an anyon $a$ by $q_a$.

Now suppose dragging a monopole from the outside of a TCI, which is assumed to be a trivial vacuum, to its inside. A $2\pi$ flux will be left on the surface after this event. Because this is a local process, no anyon far away from the flux should be able to tell the difference before and after the monopole goes through the surface. However, for an anyon with charge $q_a$, no matter how far it is from the $2\pi$ flux, it will experience an Aharonov-Bohm phase factor when it moves around the flux:
$$
\exp\left(2\pi i q_a\right)
$$
Therefore, when the monopole goes through, the $2\pi$ flux must nucleate an anyon, ${M}$, which has braiding that precisely cancels this Aharonov-Bohm phase factor for any anyon $a$. That is, there must be an anyon ${M}$ such that
\beq
\theta_{{M},a}=\exp\left(-2\pi i q_a\right)
\eeq
for all anyon $a$, where $\theta_{{M},a}$ denotes the braiding phase factor between ${M}$ and $a$. Because ${M}$ has Abelian braiding with all anyons, it must be an Abelian anyon. \cite{Barkeshli2014}

Therefore, this anyon ${M}$ is a surface avatar of the bulk monopole, and from how ${M}$ transforms under the relevant symmetries we can read off the quantum numbers of the monopole under these symmetries.

Before finishing this section and applying these methods to study some specific TCIs, let us make some remarks. All the TCIs studied in this paper have a free-fermion realization, so the first method based on symmetric gapless surface is applicable to all of them. For TCIs that cannot be realized by free fermions (see, for example, Ref. \onlinecite{Song2016}), it will be difficult to apply the first method. If such states have a symmetric STO, then the second method also applies. Notice we will describe examples where the surface exhibits the phenomenon of symmetry enforced gaplessness first discussed in Ref. \onlinecite{Wang2014}: as long as the surface preserves all the relevant symmetries, it must be gapless and a symmetric STO is disallowed. Our examples are the first crystalline-symmetric realizations of this phenomenon. Although we do not know any example so far, there can in principle be intrinsically interacting TCIs that exhibit symmetry-enforced-gapless surfaces. Such TCIs will be very interesting, and it will be difficult to use either method to read off the quantum numbers of their monopoles.

\section{Examples} \label{sec: examples}

In this section we study several examples of TCIs, and we will use the methods described above to show that they all have nontrivial monopoles, so they are all stable under interactions.

\subsection{Weak topological insulators} \label{subsec: weak-TI}

We start from a weak topological insulator (WTI), which is a TCI protected by a $U(1)$ charge conservation, a time reversal, $\mc{T}$, and a translation, $T_y$. A WTI can be smoothly connected to layers of decoupled two dimensional topological insulators.\cite{Hasan2010,Qi2010,Hasan2010a} Both a symmetric gapless surface and a symmetric STO of a WTI have been discussed in Ref. \onlinecite{Mross2015}. Below we will derive the quantum numbers of the monopole of a WTI based on these surfaces.

Let us start from the symmetric gapless surface on the $z=0$ plane, which can be described by two Dirac cones at low energies:
\beq \label{eq: Dirac-cones}
H=\psi^\dag(-i\partial_x\sigma_x-i\partial_y\sigma_z)\psi
\eeq
where $\psi=(\psi_1,\psi_2)^T$ collectively denotes the two Dirac cones, $\psi_1$ and $\psi_2$. The $\sigma$ matrices are Pauli matrices acting within each Dirac cone, and we will use $\tau$ matrices to denote Pauli matrices acting between $\psi_1$ and $\psi_2$. These will be the notations throughout this paper.

In order to determine the quantum numbers of a monopole of a WTI, let us imagine threading a $2\pi$ flux through the surface Dirac cones described by (\ref{eq: Dirac-cones}). In this flux background, the Dirac cones $\psi_1$ and $\psi_2$ will contribute a zero mode $f_1$ and $f_2$, respectively. Let us denote $|0\ra$ to be the state under this $2\pi$ flux background with no zero modes being occupied, so $f_{1,2}^\dag|0\ra$ is the state under the flux background with $f_{1,2}$ being occupied, and $f_1^\dag f_2^\dag|0\ra$ is the state under the flux background with both zero modes being occupied. To proceed, let us momentarily assume the system also has an extra anti-unitary particle-hole symmetry that flips the charge but keeps the flux. Under this assumption, the states $f_1^\dag|0\ra$ and $f_2^\dag|0\ra$ correspond to two components of the neutral monopole operator.\cite{Borokhov2002} By abuse of notations, we will denote these two components of the monopole operators by ${M}_{1,2}$ such that the correspondence is
\beq \label{eq: state-operator-correspondence}
{M}_{1}\sim f_{1}^\dag|0\ra,
\quad
{M}_{2}\sim f_{2}^\dag|0\ra
\eeq
We will determine the quantum numbers of the monopole in the presence of this extra symmetry, and then break this extra symmetry. Due to the discrete nature of the nontrivial quantum numbers, they are rigid characterizations of the monopoles which should not change when the extra symmetry is broken (at least weakly).

Suppose the translation symmetry responsible for the symmetry protection is $T_y$, translation along the $y$ direction, then the two Dirac cones have distinct momenta along the $y$ direction that differ by $\pi$ (the translation unit in the $y$ direction is taken to be $1$).\cite{Mross2015} Therefore, the symmetry actions on the low energy theory can be written as
\beq \label{eq: WTI-symmetries}
\begin{split}
&U(1): \psi\rightarrow e^{i\theta}\psi\\
&\mc{T}: \psi\rightarrow i\sigma_y\psi\\
&T_y: \psi\rightarrow \tau_z\psi
\end{split}
\eeq
Given these symmetries, there are two types of quantum numbers that the monopoles can carry: $\mc{T}$ and $T_y$ can either commute or anti-commute. The former is a set of trivial quantum numbers, and the latter corresponds to a set of nontrivial projective quantum numbers.

Now we just need to examine how $\mc{T}$ and $T_y$ act on $f_1^\dag|0\ra$ and $f_2^\dag|0\ra$. It is straightforward to examine the action of $T_y$:
\beq \label{eq: WTI-Ty}
\begin{split}
&M_1\sim f_1^\dag|0\ra\rightarrow f_1^\dag|0\ra\sim{M}_1\\
&{M}_2\sim f_2^\dag|0\ra\rightarrow -f_2^\dag|0\ra\sim-{M}_2
\end{split}
\eeq
Extra care is needed when examining the action of $\mc{T}$, because this operation will change the $2\pi$ flux background into a $-2\pi$ flux background, which has two other zero modes $\tilde f_{1,2}$ contributed by $\psi_{1,2}$. Denote $|\tilde 0\ra$ to be the state under a $-2\pi$ flux background with no zero mode being occupied, then $\tilde f_{1,2}^\dag|\tilde 0\ra$ is the state under the $-2\pi$ flux background with one of the zero modes being occupied, and $\tilde f_1^\dag \tilde f_2^\dag|\tilde 0\ra$ is the state under the $-2\pi$ flux background with both zero modes being occupied. Then under $\mc{T}$
\beq
\begin{split}
&{M}_1\sim f_1^\dag|0\ra\rightarrow\tilde f_1^\dag|\tilde 0\ra\\
&{M}_2\sim f_2^\dag|0\ra\rightarrow\tilde f_2^\dag|\tilde 0\ra
\end{split}
\eeq
Ref. \onlinecite{Zou2017} (see Appendix F 4 therein) showed that under the state-operator-correspondence (\ref{eq: state-operator-correspondence}),
\beq
{M}_1^\dag\sim\tilde f_2^\dag|\tilde 0\ra,
\quad
{M}_2^\dag\sim-\tilde f_1^\dag|\tilde 0\ra
\eeq
up to an unimportant phase factor. So the above equation means under $\mc{T}$
\beq \label{eq: WTI-T}
{M}_1\rightarrow-{M}_2^\dag,
\quad
{M}_2\rightarrow{M}_1^\dag
\eeq

Now combining (\ref{eq: WTI-Ty}) and (\ref{eq: WTI-T}), we see the actions of $T_y$ and $\mc{T}$ anti-commute on the monopoles, which represents a nontrivial projective representation of the original symmetry described in (\ref{eq: WTI-symmetries}). Therefore, we conclude that the monopoles in a WTI carry nontrivial quantum numbers, and a WTI is stable under interactions as long as the relevant symmetry are preserved.

As a consistency check, let us rederive the quantum numbers of the monopole by studying a symmetric STO of a WTI. As discussed in Ref. \onlinecite{Mross2015}, a WTI allows a symmetric $Z_4$ STO, where the anyon contents can be written as
\beq \label{eq: Z4-STO}
\{1,e,m,e^\alpha m^\beta\}\times\{1,\psi\}
\eeq
where $e$ can be viewed as the elementary $Z_4$ charge, $m$ can be viewed as the elementary $Z_4$ flux, such that $\theta_{e,m}=i$, and $\psi$ is a local fermion. $\alpha$ and $\beta$ takes integer values (mod 4).

In this $Z_4$ STO, $e$ carries charge-1/2 and $m$ is neutral. Based on the principle in Sec. \ref{subsec: monopole-STO}, $m^2$ is the surface avatar of the bulk monopole. According to the symmetry assignments of this STO in Ref. \onlinecite{Mross2015}, it is straightforward to check that $\mc{T}$ and $T_y$ anti-commute on $m^2$, which means these two symmetries indeed anti-commute on the monopole. This is consistent with the above result obtained from the symmetric gapless surface, and it confirms that a WTI is stable under interactions as long as the symmetry in (\ref{eq: WTI-symmetries}) is preserved.

\subsection{Topological crystalline insulators protected by a mirror symmetry} \label{subsec: mirror-TCI}

Next we discuss TCIs protected by a $U(1)$ charge conservation and a mirror symmetry $\mc{M}$ with respect to the $x=0$ plane, to which class the experimentally discovered TCI SnTe belongs. We denote such a TCI by MTCI.

We will focus on the case where the surface can host two Dirac fermions, whose low energy Hamiltonian is still given by (\ref{eq: Dirac-cones}). In this case, the symmetry assignment is
\beq \label{eq: MTCI-symmetries}
\begin{split}
&U(1): \psi\rightarrow e^{i\theta}\psi\\
&\mc{M}: \psi(x,y)\rightarrow\sigma_x\psi(-x,y)
\end{split}
\eeq
where we have assumed this surface is on the $z=0$ plane. For these symmetries, $\mc{M}^2$ can be $\pm 1$ on the monopole, and that $\mc{M}^2=1$ corresponds to trivial quantum numbers and $\mc{M}^2=-1$ corresponds to nontrivial projective quantum numbers.

To determine how $\mc{M}$ acts on the monopoles, we only need to examine the action of $\mc{M}$ on $f_{1,2}^\dag|0\ra$. Notice under the reflection a $2\pi$ flux also needs to be converted into a $-2\pi$ flux, and it is straightforward to check that under $\mc{M}$
\beq \label{eq: MTCI-symmetries-M}
\begin{split}
&M_1\sim f_1^\dag|0\ra\rightarrow \tilde f_1^\dag|\tilde 0\ra\sim-M_2^\dag\\
&M_2\sim f_2^\dag|0\ra\rightarrow \tilde f_2^\dag|\tilde 0\ra\sim M_1^\dag
\end{split}
\eeq
Therefore, $\mc{M}$ squares to $-1$ on the monopole. This is a nontrivial projective representation of the symmetry (\ref{eq: MTCI-symmetries}), so this TCI is stable under interactions as long as the symmetry (\ref{eq: MTCI-symmetries}) is preserved.

As a consistency check, we can also derive the quantum numbers of the monopole from a symmetric STO of this TCI. This TCI can have a symmetric $Z_4$ STO similar to the one in (\ref{eq: Z4-STO})\cite{Cheng2017,Hong2017}. In particular, in this case $e$ carries charge-1/2 and $m$ is neutral, too. Then based on the principle in Sec. \ref{subsec: monopole-STO}, $m^2$ is the surface avatar of the bulk monopole. As shown in Ref. \onlinecite{Cheng2017,Hong2017} (see also Appendix \ref{app: STO-double-FKM}), indeed, $\mc{M}^2=-1$ on $m^2$. This is consistent with the above result obtained by studying the symmetric gapless surface.

This TCI is compatible with a further time reversal symmetry, and we discuss two time reversal symmetric generalizations of this state in Appendix \ref{app: generalized-mirror-TCI}. These two states are denoted as TMTCI-1 and TMTCI-2.

\subsection{Doubled topological insulators protected by a mirror symmetry} \label{subsec: double-Fu-Kane-Mele}

As our third example, we consider two copies of topological insulators further equipped with a mirror symmetry (DTI). It is well known that a single topological insulator is protected by $U(1)$ charge conservation and time reversal, but two copies of them are smoothly connected to a trivial insulator if there is no other symmetry.\cite{Hasan2010,Qi2010,Hasan2010a} However, with a further mirror symmetry $\mc{M}$, two copies of topological insulators can still be nontrivial at the free-fermion level. Such an insulator still has a gapless surface described by (\ref{eq: Dirac-cones}), and the symmetries are assigned as
\beq \label{eq: double-FKM-symmetries}
\begin{split}
&U(1): \psi\rightarrow e^{i\theta}\psi\\
&\mc{T}: \psi\rightarrow i\sigma_y\psi\\
&\mc{M}: \psi(x,y)\rightarrow \sigma_z\tau_y\psi(-x,y)
\end{split}
\eeq
It is straightforward to check that no fermion bilinear term can be written down to fully gap out (\ref{eq: Dirac-cones}) while preserving all symmetries.

To understand whether this state is stable under interactions, let us derive the quantum numbers of the monopole from the above surface state. For the above symmetries, the nontrivial projective quantum numbers on monopoles correspond to $\mc{M}^2=-1$ or $Z^2\equiv(\mc{TM})^2=-1$.

Because the $\mc{T}$ action in this state is the same as in a WTI, under $\mc{T}$ we again have (\ref{eq: WTI-T}). It is straightforward to check that under $\mc{M}$
\beq
\begin{split}
&M_1\sim f_1^\dag|0\ra\rightarrow -i\tilde f_2^\dag|\tilde 0\ra\sim -iM_1^\dag\\
&M_2\sim f_2^\dag|0\ra\rightarrow i\tilde f_1^\dag|\tilde 0\ra\sim -iM_2^\dag
\end{split}
\eeq
So under $Z\equiv\mc{TM}$, the product of $\mc{T}$ and $\mc{M}$,
\beq
M_1\rightarrow iM_2,
\quad
M_2\rightarrow -iM_1
\eeq
Therefore, $\mc{M}$ squares to $1$ and $Z\equiv\mc{TM}$ squares to $-1$ on the monopole. So the monopole indeed carries nontrivial quantum numbers, and thus this insulator is stable under interactions.

We can also derive these quantum numbers of the monopoles by examining the symmetric STO of this TCI. As shown in Appendix \ref{app: STO-double-FKM}, this TCI can have a symmetric $Z_4$ STO similar to the ones discussed earlier, and we find consistent results for the quantum numbers of monopoles with the above.

Notice both $\mc{T}$ and $\mc{M}$ are important for the symmetry protection of this TCI, and as long as one of them is broken, a symmetric fermion bilinear term can be written down to fully gap the surface. In fact, in Appendix \ref{app: STO-double-FKM} we argue that a DTI can be obtained from a WTI by suitably breaking the protecting translation symmetry of the latter. In Appendix \ref{app: generalized-mirror-TCI}, we will discuss TMTCI-2, another example which can also be viewed as two copies of topological insulators protected by a further mirror symmetry. However, in TMTCI-2 the mirror symmetry alone is sufficient for the symmetry protection. In fact, that state is just a time reversal symmetric version of the state MTCI described in Sec. \ref{subsec: mirror-TCI}.

\subsection{Topological crystalline insulators with symmetry-enforced-gapless surfaces} \label{subsec: SEG}

As our final examples, we present two TCIs that exhibit symmetry-enforced-gapless surfaces. Both TCIs have a symmetric gapless surface state described by (\ref{eq: Dirac-cones}) at low energies, and they differ by the symmetries.

The first such TCI, SEG-1, has a $U(1)$ charge conservation, a mirror symmetry $\mc{M}$ and an $SU(2)$ symmetry. These symmetries are assigned as
\beq \label{eq: FTP-symmetries}
\begin{split}
&U(1): \psi\rightarrow e^{i\theta}\psi\\
&\mc{M}: \psi(x,y)\rightarrow\sigma_z\psi(-x,y)\\
&SU(2): \psi\rightarrow U_\tau\psi
\end{split}
\eeq
where $U_\tau$ means an $SU(2)$ matrix generated by $\tau$'s. For these symmetries, there are three types of nontrivial projective quantum numbers on monopoles: having $\mc{M}^2=-1$ and/or having spin-1/2.

It is easy to see this is just an $SU(2)$ symmetric version of MTCI discussed in Sec. \ref{subsec: mirror-TCI}, so $\mc{M}$ squares to $-1$ on the monopole. Furthermore, it is straightforward to check that the monopole also transforms as a spin-1/2 under the $SU(2)$ symmetry. Therefore, the monopole is nontrivial, and this TCI is stable under interactions.

Now we show by contradiction that this TCI does not allow a symmetric STO. Suppose this TCI can have a symmetric STO, then the anyons of this STO have definite spins under the $SU(2)$ symmetry. Because the local fermion carries spin-1/2 under the $SU(2)$, it is always possible to make the an anyon a spin-singlet by attaching certain local fermions to it. Then the STO can be written as
\beq
\{1,a,\cdots\}\times\{1,\psi\}
\eeq
where the sector $\{1,a,\cdots\}$ includes all topologically nontrivial spinless anyons, and the sector $\{1,\psi\}$ includes the spin-1/2 local fermion. The first sector is already closed under fusion and braiding due to the $SU(2)$ symmetry. Moreover, these two sectors will not be mixed under the symmetry actions, because the symmetries cannot change a spin-singlet into a spin-1/2, or vice versa. Therefore, the sector $\{1,a,\cdots\}$ alone must be able to emerge from a system made of local spin-singlets. In this system, all local spin-singlets must be bosons that carry even $U(1)$ charges. Notice the symmetries of these bosons include a $U(1)$ charge conservation and a mirror symmetry $\mc{M}$.

Now consider the possible properties of the elementary monopoles of a system made of these bosons. Because the charges of such bosons are even, the elementary monopoles of these bosons only emit $\pi$ flux. That is to say, two of such monopoles should be the $2\pi$ monopole discussed earlier, which carries spin-1/2 and $\mc{M}^2=-1$. Based on the understanding of a bosonic TCI in general, these ``half monopoles" can be either a boson or a fermion, either a spin-singlet or a spin-1/2, and $\mc{M}$ can square to either $1$ or $-1$. However, in none of these cases the $2\pi$ monopole will be a spin-1/2 boson with $\mc{M}^2=-1$. This contradiction shows that this TCI cannot have a symmetric STO, and its surface is enforced to be gapless as long as the symmetries (\ref{eq: FTP-symmetries}) are preserved. This phenomenon of the surface is dubbed ``symmetry enforced gaplessness",\cite{Wang2014} and our example provides a crystalline-symmetric realization of it.

Now we turn to SEG-2, the second TCI that has symmetry-enforced-gapless surface. This TCI has a $U(1)$ charge conservation, an $SU(2)$ symmetry and a symmetry $\mc{P}$ that can be viewed as a combination of a mirror symmetry and a unitary charge conjugation. The low energy Hamiltonian of the surface is still described by (\ref{eq: Dirac-cones}), while the symmetries are assigned as
\beq
\begin{split}
&U(1): \psi\rightarrow e^{i\theta}\psi\\
&SU(2): \psi\rightarrow U_\tau\psi\\
&\mc{P}: \psi(x,y)\rightarrow\sigma_z\tau_y\psi^*(-x,y)
\end{split}
\eeq
It is easy to check that no fermion bilinear term can be written down to fully gap (\ref{eq: Dirac-cones}) without breaking these symmetries, so this is a nontrivial TCI at the level of free fermions. In addition, it is straightforward to see that the monopole of this TCI carries nontrivial projective quantum number, {\it i.e.} spin-1/2 under the $SU(2)$ symmetry, so this nontrivial TCI is stable under interactions. Furthermore, a similar argument as above shows that this TCI also allows no symmetric STO and its surface exhibits symmetry enforced gaplessness.

\section{Relation to symmetry enriched $U(1)$ quantum spin liquids} \label{sec: relation to SEU(1)QSL}

After understanding the properties of the monopoles of these TCIs in details, in this section we discuss the relations between these TCIs and symmetry enriched $U(1)$ QSLs. To this end, let us first briefly review the physics of symmetry enriched $U(1)$ QSLs.\cite{Zou2017}

A three dimensional $U(1)$ QSL is a three dimensional spin system that has emergent gapless photons at low energies. In a condensed matter system, the appearance of an emergent gapless photon is necessarily associated with the existence of emergent electric charges and magnetic monopoles (and their bound states, dyons). These fractional excitations can in principle also be gapless, but we assume only the photons are gapless for simplicity. The existence of fractional excitations implies long-range entanglement in the ground state, and this nontrivial nature is independent of any symmetry.

In the absence of any symmetry, all $U(1)$ QSLs can be smoothly connected to each other. However, when there are global symmetries in $U(1)$ QSLs, there can be different symmetry enriched $U(1)$ QSLs due to symmetry protected distinctions. In many cases, their differences are reflected in their different spectra. Namely, in different symmetry enriched $U(1)$ QSLs, their electric charges and magnetic monopoles can have different statistics and/or quantum numbers.

It turns out very helpful to view $U(1)$ QSLs as gauged insulators, {\it i.e.} an insulator coupled with a dynamical $U(1)$ gauge field. The properties of the insulator will then determine the statistics and quantum numbers of both the electric charge and the magnetic monopole, thus determine the property of the symmetric $U(1)$ QSL.

Therefore, we will consider coupling the TCIs discussed above to a dynamical $U(1)$ gauge field. This gauging procedure turns the TCIs into a $U(1)$ QSLs with some global symmetries. Below we will discuss our examples in turn.

Let us start with a WTI. After gauging it becomes a $U(1)$ QSL with time reversal $\mc{T}$ and translation $T_y$. In this $U(1)$ QSL, the electric charge is a fermion. This fermionic charge is a Kramers doublet under the original time reversal $\mc{T}$, and also a ``Kramers doublet" under a new anti-unitary symmetry $\mc{T}'$, which is generated by the product of the generators of $\mc{T}$ and $T_y$. As discussed earlier, the magnetic monopole is a boson, and $\mc{T}$ and $T_y$ anti-commute on the monopole. For these reasons, we denote this $U(1)$ QSL by $E_{fTT'}M_{b-}$.

Next we turn to MTCI discussed in Sec. \ref{subsec: mirror-TCI}. After gauging, the symmetry of the resulting $U(1)$ QSL is just a mirror symmetry $\mc{M}$. For later purposes, it is more convenient to twist our notations here. That is, we will regard the fermion of the TCI as the magnetic monopole of the resulting $U(1)$ QSL, which then identifies the monopole of the TCI discussed in Sec. \ref{subsec: mirror-TCI} as the electric charge of this $U(1)$ QSL. So the magnetic monopole is a fermion, and the electric charge is a boson that has $\mc{M}^2=-1$. For this reason, we denote this $U(1)$ QSL by $E_{bM}M_{f}$.

Now we turn to the DTI discussed in Sec. \ref{subsec: double-Fu-Kane-Mele}. After gauging the symmetry of the resulting $U(1)$ QSL is $\mc{T}\times\mc{M}$, where both $\mc{T}$ and $\mc{M}$ square to $1$ for all local excitations. The electric charge of the $U(1)$ QSL will be taken as the fermion that is a Kramers doublet under $\mc{T}$. It is also easy to see $Z^2\equiv(\mc{TM})^2=-1$ for the charge. The magnetic monopole has $Z^2=-1$. For this reason, we denote this $U(1)$ QSL by $(E_{fTZ}M_{bZ})_-$, with the parenthesis and minus sign implying that the system has both a time reversal and a mirror symmetry, and the mirror symmetry action preserves the electric charge.

Finally we consider the two TCIs that exhibit symmetry-enforced-gapless surfaces discussed in Sec. \ref{subsec: SEG}. For SEG-1, after gauging the symmetry becomes $SO(3)\times\mc{M}$, which means all local excitations carry integer spins and $\mc{M}^2=1$. In this case, it is again more convenient to identify the fermion of the TCI as the magnetic monopole of the resulting $U(1)$ QSL, and this fermion is a spin-1/2. This lets us identify the bosonic magnetic monopole of the TCI as the electric charge of the $U(1)$ QSL, which is a spin-1/2 boson that has $\mc{M}^2=-1$. For this reason, we denote it by $E_{bM\half}M_{f\half}$. As for SEG-2, the symmetry of the resulting $U(1)$ QSL is $SO(3)\times\mc{P}$, which means all local excitations have integer spins and $\mc{P}^2=1$. Notice after gauging $\mc{P}$ should be interpreted as a mirror symmetry in terms of the local excitations, so from now on we will denote $\mc{P}$ as $\mc{M}$ for notational consistency. We will identify the electric charge of the resulting $U(1)$ QSL to be the fermion of the TCI, which carries spin-1/2 and has $\mc{M}^2=-1$. Then the magnetic monopole of the $U(1)$ QSL is identified as the magnetic monopole of the TCI, which is a spin-1/2 boson. For this reason, we denote this $U(1)$ QSL by $E_{fM\half}M_{b\half}$.

The above identification of the gauged TCIs discussed here and some crystalline symmetric $U(1)$ QSLs are summarized in Table \ref{table: TCI-QSL}.

\begin{widetext}
\begin{table*}
\centering
\begin{tabular}{c|c|c|c|c|c|c|c}
\hline
 & WTI & MTCI & DTI & SEG-1 & SEG-2 & TMTCI-1 & TMTCI-2\\
\hline
section in the paper & \ref{subsec: weak-TI} & \ref{subsec: mirror-TCI} & \ref{subsec: double-Fu-Kane-Mele} & \ref{subsec: SEG} & \ref{subsec: SEG} & \ref{app: sub-singlet} & \ref{app: sub-doublet}\\
\hline
$U(1)$ QSL & $E_{fTT'}M_{b-}$ & $E_{bM}M_f$ & $(E_{fTZ}M_{bZ})_-$ & $E_{bM\half}M_{f\half}$ & $E_{fM\half}M_{b\half}$ & $(E_{fZ}M_{bMZ})_-$ & $(E_{fT}M_{bM})_-$\\
\hline
symmetry & $\mc{T}\times T_y$ & $\mc{M}$ & $\mc{T}\times\mc{M}$ & $SO(3)\times\mc{M}$ & $SO(3)\times\mc{M}$ & $\mc{T}\times\mc{M}$ & $\mc{T}\times\mc{M}$\\
\hline
Analog with internal symmetries & $E_{fTT'}M_{b-}$ & $E_{bT}M_f$ & $(E_{fTZ}M_{bZ})_-$ & $E_{bT\half}M_{f\half}$ & $E_{f\half}M_{b\half}$ & $(E_{fZ}M_{bT'Z})_-$ & $(E_{fT}M_{bT'})_-$\\
\hline
symmetry & $\mc{T}\times Z_2$ & $\mc{T}$ & $\mc{T}\times\mc{T}'$ & $SO(3)\times\mc{T}$ & $SO(3)\times\mc{T}$ & $\mc{T}\times\mc{T}'$ & $\mc{T}\times\mc{T}'$\\
\hline
\end{tabular}
\caption{Relation between the TCIs and symmetry enriched $U(1)$ QSLs. The first row lists the TCIs we consider, which are discussed in sections and appendices specified in the second row. The third row lists the corresponding $U(1)$ QSLs obtained by coupling these TCIs to a dynamical $U(1)$ gauge field, and the symmetries of these $U(1)$ QSLs are listed in the fourth row. The fifth row lists the analogs of these crystalline symmetric $U(1)$ QSLs realized with internal symmetries, whose symmetries are specified in the last row. The detailed properties of these $U(1)$ QSLs are described in the main text and in Table \ref{table: TTy} - Table \ref{table: SO(3)T-U(1)}, which clearly illustrate the correspondence between the crystalline symmetric $U(1)$ QSLs and internal symmetric $U(1)$ QSLs.} \label{table: TCI-QSL}
\end{table*}
\end{widetext}

There is actually another level of relation between the TCIs and the symmetry enriched $U(1)$ QSLs: the $U(1)$ QSLs enriched by crystalline symmetries have their analogs with internal symmetries discussed in Ref. \onlinecite{Zou2017,Wang2015} (see Table \ref{table: TCI-QSL}). Such a correspondence was first noted in Ref. \onlinecite{Song2016} and later elaborated in Ref. \onlinecite{Thorngren2016} as the ``crystalline equivalence principle". Below we illustrate this correspondence.{\footnote{The $U(1)$ QSLs with internal symmetries can also be viewed as gauged insulators, and there is also a correspondence between the TCIs studied in this paper and those insulators. However, in this paper we focus on the correspondence between the $U(1)$ QSLs with crystalline symmetries and those with internal symmetries.}}

Let us start with the WTI. Notice the translation symmetry $T_y$ of a WTI acts as an internal $Z_2$ symmetry at low energies, so we can ask if there is a $\mc{T}\times Z_2$ symmetric $U(1)$ QSL that has analogous properties as $E_{fTT'}M_{b-}$, the gauged WTI. As shown in Ref. \onlinecite{Zou2017}, there is indeed a $\mc{T}\times Z_2$ symmetric $U(1)$ QSL also dubbed $E_{fTT'}M_{b-}$. The properties of these two states are listed in Table \ref{table: TTy}, and the correspondence is clear.

For other crystalline-symmetric $U(1)$ QSLs that involve a mirror symmetry $\mc{M}$, if we replace the mirror symmetry $\mc{M}$ by a time reversal symmetry $\mc{T}'$, we can find their corresponding states in Ref. \onlinecite{Zou2017} as well. However, we need to pay extra care when carrying out this procedure: if the charge (monopole) of the $U(1)$ QSL is a fermion and $\mc{M}$ acts on the charge as a mirror reflection combined with a unitary charge conjugation, then on the charge (monopole) $\mc{M}^2=\pm 1$ corresponds to $\mc{T}'^2=\mp 1$.\cite{Song2016}

Now we demonstrate this more explicitly. The $\mc{M}$ symmetric $E_{bM}M_f$ corresponds to the $\mc{T}$ symmetric $E_{bT}M_f$, where the electric charge has $\mc{T}^2=-1$ for the electric charge (see Table \ref{table: M}). The $\mc{T}\times\mc{M}$ symmetric $(E_{fTZ}M_{bZ})_-$, $(E_{fZ}M_{bMZ})_-$ and $(E_{fT}M_{bM})_-$ correspond to the $\mc{T}\times\mc{T}'$ symmetric $(E_{fTZ}M_{bZ})_-$, $(E_{fZ}M_{bT'Z})_-$ and $(E_{fT}M_{bT'})_-$, respectively. The symmetries in the three latter states can be more conveniently phrased as a time reversal $\mc{T}$ and a unitary $Z_2$ charge conjugation $\mc{C}$, which is generated by the product of the generators of $\mc{T}$ and $\mc{T}'$. The properties of these states are listed in Table \ref{table: Z2Z2T-U(1)}, and the correspondence is clear. The $SO(3)\times\mc{M}$ symmetric $E_{bM\half}M_{f\half}$ and $E_{fM\half}M_{b\half}$ correspond to the $SO(3)\times\mc{T}$ symmetric $E_{bT\half}M_{f\half}$ and $E_{f\half}M_{b\half}$, respectively. The properties of these states are summarized in Table \ref{table: SO(3)T-U(1)}, and one can again see the clear correspondence. Ref. \onlinecite{Zou2017} showed that the latter two states possess fractional topological responses to an external $SO(3)$ gauge field, so they are dubbed ``fractional topological paramagnets". $E_{bM\half}M_{f\half}$ and $E_{fM\half}M_{b\half}$ are the crystalline-symmetric realizations of fractional topological paramagnets.

\begin{table}
\centering
\begin{tabular}{c|c|c|c}
\hline
$\mc{T}\times T_y$ & $\mc{T}_E^2$ & $\mc{T'}_E^2$ & $[\mc{T},T_y]_M$\\
\hline
$E_{fTT'}M_{b-}$ & $-1$ & $-1$ & $-$\\
\hline
\end{tabular}

\begin{tabular}{c|c|c|c}
\hline
$\mc{T}\times Z_2$ & $\mc{T}_E^2$ & $\mc{T'}_E^2$ & $[\mc{T},Z_2]_M$\\
\hline
$E_{fTT'}M_{b-}$ & $-1$ & $-1$ & $-$\\
\hline
\end{tabular}
\caption{Upper: properties of the $\mc{T}\times T_y$ symmetric $E_{fTT'}M_{b-}$. $\mc{T}_E^2$ stands for the value of $\mc{T}^2$ on the electric charge, $\mc{T'}_E^2$ stands for the value of $\mc{T'}^2\equiv (\mc{T}T_y)^2$ on the charge, and $[\mc{T},T_y]_M=+$ and $[\mc{T},T_y]_M=-$ mean that $\mc{T}$ and $T_y$ commute or anti-commute on the monopole, respectively. Lower: properties of the the $\mc{T}\times Z_2$ symmetric $E_{fTT'}M_{b-}$. $\mc{T}_E^2$ stands for the value of $\mc{T}^2$ on the electric charge, $\mc{T'}_E^2$ stands for the value of $\mc{T'}^2\equiv (\mc{T}Z_2)^2$ on the charge, and $[\mc{T},Z_2]_M=+$ and $[\mc{T},Z_2]_M=-$ mean that $\mc{T}$ and $Z_2$ commute or anti-commute on the monopole, respectively.} \label{table: TTy}
\end{table}

\begin{table}
\centering
\begin{tabular}{c|c}
\hline
$\mc{M}$ & $\mc{M}_E^2$\\
\hline
$E_{bM}M_f$ & $-1$\\
\hline
\end{tabular}

\begin{tabular}{c|c}
\hline
$\mc{T}$ & $\mc{T}_E^2$\\
\hline
$E_{bT}M_f$ & $-1$\\
\hline
\end{tabular}
\caption{Upper: properties of the $\mc{M}$ symmetric $E_{bM}M_f$, where $\mc{M}_E^2$ stands for the value of $\mc{M}^2$ on the electric charge. Lower: properties of the $\mc{T}$ $E_{bT}M_f$, where $\mc{T}_E^2$ stands for the value of $\mc{T}^2$ on the electric charge.} \label{table: M}
\end{table}

\begin{table}[h!]
\centering
\begin{tabular}{c|c|c|c|c}
\hline
$\mc{T}\times\mc{M}$ & $\mc{T}_E^2$ & $Z_E^2$ & $\mc{M}_M^2$ & $Z_M^2$\\
\hline
$(E_{fTZ}M_{bZ})_-$ & $-1$ & $-1$ & $1$ & $-1$\\
\hline
$(E_{fZ}M_{bMZ})_-$ & $1$ & $-1$ & $-1$ & $-1$\\
\hline
$(E_{fT}M_{bM})_-$ & $-1$ & $1$ & $-1$ & $1$\\
\hline
\end{tabular}

\begin{tabular}{c|c|c|c|c}
\hline
$\mc{T}\times\mc{T}'$ & $\mc{T}_E^2$ & $\mc{C}_E^2$ & $\mc{T'}_M^2$ & $\mc{C}_M^2$\\
\hline
$(E_{fTZ}M_{bZ})_-$ & $-1$ & $-1$ & $1$ & $-1$\\
\hline
$(E_{fZ}M_{bT'Z})_-$ & $1$ & $-1$ & $-1$ & $-1$\\
\hline
$(E_{fT}M_{bT'})_-$ & $-1$ & $1$ & $-1$ & $1$\\
\hline
\end{tabular}
\caption{Upper: properties of the $\mc{T}\times\mc{M}$ symmetric $(E_{fTZ}M_{bZ})_-$, $(E_{fZ}M_{bMZ})_-$ and $(E_{fT}M_{bM})_-$. $\mc{T}_E^2$ stands for the value of $\mc{T}^2$ on the electric charge, $\mc{M}_M^2$ stands for the value of $\mc{M}^2$ on the magnetic monopole, and $Z_E^2$ and $Z_M^2$ stand for the value of $Z^2\equiv(\mc{TM})^2$ on the charge and monopole, respectively. Lower: properties of the $\mc{T}\times\mc{T}'$ symmetric $(E_{fTZ}M_{bZ})_-$, $(E_{fZ}M_{bT'Z})_-$ and $(E_{fT}M_{bT'})_-$. $\mc{T}_E^2$ stands for the value of $\mc{T}^2$ on the electric charge, $\mc{T'}_M^2$ stands for the value of $\mc{T'}^2$ on the magnetic monopole, and $\mc{C}_{E}^2$ and $\mc{C}_M^2$ stand for the value of $\mc{C}^2$ on the charge and the monopole, respectively.} \label{table: Z2Z2T-U(1)}
\end{table}

\begin{table}[h!]
\centering
\begin{tabular}{c|c|c|c}
\hline
$SO(3)\times\mc{M}$ & $\mc{M}_E^2$ & $S_E$ & $S_M$\\
\hline
$E_{bM\half}M_{f\half}$ & $-1$ & $\half$ & $\half$\\
\hline
$E_{fM\half}M_{b\half}$ & $-1$ & $\half$ & $\half$\\
\hline
\end{tabular}

\begin{tabular}{c|c|c|c}
\hline
$SO(3)\times\mc{T}$ & $\mc{T}_E^2$ & $S_E$ & $S_M$\\
\hline
$E_{bT\half}M_{f\half}$ & $-1$ & $\half$ & $\half$\\
\hline
$E_{f\half}M_{b\half}$ & $1$ & $\half$ & $\half$\\
\hline
\end{tabular}
\caption{Upper: properties of the $SO(3)\times\mc{M}$ symmetric $U(1)$ QSLs $E_{bM\half}M_{f\half}$ and $E_{fM\half}M_{b\half}$. $\mc{M}_E^2$ stands for the value of $\mc{M}^2$ on the electric charge, and $S_E$ and $S_M$ stand for the spin of the charge and the monopole, respectively. Lower: properties of the $SO(3)\times\mc{T}$ symmetric $U(1)$ QSLs $E_{bT\half}M_{f\half}$ and $E_{f\half}M_{b\half}$. $\mc{T}_E^2$ stands for the value of $\mc{T}^2$ on the electric charge, and $S_E$ and $S_M$ stand for the spin of the charge and the monopole, respectively.} \label{table: SO(3)T-U(1)}
\end{table}

The above relation may provide simpler ways to study difficult problems on one side of the correspondence, by translating it to a possibly simpler problem on the other side. For example, if the above correspondence is generally true, we expect the classification and characterization of some crystalline symmetric $U(1)$ QSLs can be directly read off from Ref. \onlinecite{Zou2017} (and Ref. \onlinecite{Wang2015}). More specifically, the existence of 7 different $\mc{T}$ symmetric $U(1)$ QSLs leads to the  expectation of 7 different $\mc{M}$ symmetric $U(1)$ QSLs, the existence of $15$ different $SO(3)\times\mc{T}$ symmetric $U(1)$ QSLs leads to the expectation of 15 different $SO(3)\times\mc{M}$ symmetric $U(1)$ QSLs, and the existence of $38$ different $\mc{T}\times\mc{T}'$ symmetric $U(1)$ QSLs leads to the expectation of $38$ different $\mc{T}\times\mc{M}$ symmetric $U(1)$ QSLs. Similar expectations hold for other symmetries. The physical characterizations of the crystalline symmetric $U(1)$ QSLs can also be read off from their analogs with internal symmetries from Ref. \onlinecite{Wang2015,Zou2017}, as done above.

\section{Discussion} \label{sec: discussion}

By directly studying the bulk properties of several band-theory-based nontrivial TCIs, we show all of them are stable under interactions, because they all have nontrivial bulk monopoles once they are coupled to a $U(1)$ gauge field. The properties of the monopoles are derived based on two complementary methods, which involve studying different types of surface states of the TCIs. That these methods give consistent answers is on the one hand assuring, and on the other hand implying that in many cases it is a more direct and unified way of to characterize nontrivial TCIs in terms of their bulk properties. However, we note that sometimes the boundary can carry subtle information of a topological phase that is not directly visible from the bulk.\cite{Kitaev2006}

All the TCIs we have discussed can be realized by free fermions, and two of them have symmetry-enforced-gapless surfaces. That is, even under strong interactions, their surface states must be gapless as long as the relevant symmetries are preserved. It is intriguing to find a TCI that on the one hand cannot be realized by free fermions, and on the other hand have symmetry-enforced-gapless surfaces. It will be an interesting challenge to study the properties of the monopoles of such TCIs.

The TCIs that we discuss are not only of conceptual importance, but are also of experimental significance. We expect they will either be synthesized in the near future, or have already been synthesized but remain to be further investigated.

We have discussed the relations between the TCIs and symmetry enriched $U(1)$ QSLs at two levels. First, when the TCIs are coupled to a dynamical $U(1)$ gauge field, they become $U(1)$ QSLs enriched by crystalline symmetries. We discuss the properties of these $U(1)$ QSLs in details. In particular, the two TCIs with symmetry-enforced-gapless surfaces are shown to be related to crystalline-symmetric realizations of ``fractional topological paramagnets" introduced in Ref. \onlinecite{Zou2017}. Second, we demonstrate an interesting correspondence between the $U(1)$ QSLs with crystalline symmetries and $U(1)$ QSLs with internal symmetries discussed in Ref. \onlinecite{Zou2017}. This is a manifestation of the crystalline equivalence principle first noted in Ref. \onlinecite{Song2016} and further elaborated in Ref. \onlinecite{Thorngren2016}. Such a correspondence may potentially provide a simpler way of solving a difficult problem, by relating a problem with internal symmetry to a problem with crystalline symmetry, or vice versa.

In this paper we focus on TCIs with a translation symmetry or a mirror symmetry, and it will be be interesting to obtain physical bulk characterizations for TCIs with other types of crystalline symmetries.

Besides the magnetic monopoles, there is another type of bulk characterizations that is perhaps more ubiquitous to TCIs: properties of defects of the crystalline symmetries, such as dislocations and disclinations. There has already been some work in this direction, and intriguing results have been found in many cases.\cite{Teo2010,Cho2014,Mross2015,You2016,You2016a,Thorngren2016} It will be interesting and important to obtain systematic physical characterizations of these defects, as well as their interplay with other defects such as monopoles. We leave this for future work.

\section{Ackownledgement}

I thank Meng Cheng, (Adrian) Hoi Chun Po, T. Senthil and Ashvin Vishwanath for helpful discussions. Especially, I thank Hoi Chun Po for suggesting that I should write a paper. I also thank Chong Wang and T. Senthil for a previous collaboration that enables me to solve the problems discussed in this paper. After this draft is submitted to arXiv, I became aware that Hoi Chun Po and Ashvin Vishwanath obtained similar results for a doubled topological insulator protected by an inversion symmetry. This work was supported by a US Department of Energy grant
de-sc0008739.

\appendix

\section{Time reversal symmetric generalizations of the mirror symmetric TCI} \label{app: generalized-mirror-TCI}

In this appendix we introduce two time reversal symmetric generalizations of the MTCI discussed in Sec. \ref{subsec: mirror-TCI}. Because MTCI has nontrivial monopoles and is stable under interactions even without a time reversal symmetry, this will remain the case in the presence of a further time reversal. Our main purpose is then to check if the time reversal symmetry gives further nontrivial quantum numbers to the monopoles.

The low energy surface Hamiltonian is still given by (\ref{eq: Dirac-cones}), and the $U(1)$ symmetry and $\mc{M}$ symmetry are assigned as (\ref{eq: MTCI-symmetries}). We will further equip the TCIs with a time reversal symmetry $\mc{T}$. Depending on whether the fermion is a Kramers singlet or a Kramers doublet under $\mc{T}$, we will have two different generalizations.

\subsection{Kramers singlet fermions} \label{app: sub-singlet}

First consider the $\mc{T}$ action
\beq
\mc{T}: \psi\rightarrow\sigma_y\tau_y\psi
\eeq
where the fermions are Kramers singlets under $\mc{T}$. Notice, combining this and (\ref{eq: MTCI-symmetries}), we see $Z^2\equiv(\mc{TM})^2=-1$ for the fermions. This generalization of MTCI will be labeled as TMTCI-1.

Now we check the quantum numbers of the monopoles. Under $\mc{T}$
\beq
\begin{split}
&M_1\sim f_1^\dag|0\ra\rightarrow -i\tilde f_2^\dag|\tilde 0\ra\sim-iM_1^\dag\\
&M_2\sim f_2^\dag|0\ra\rightarrow i\tilde f_1^\dag|\tilde 0\ra\sim -iM_2^\dag
\end{split}
\eeq
Combining this with (\ref{eq: MTCI-symmetries-M}) yields
\beq
M_1\rightarrow -iM_2,
\quad
M_2\rightarrow iM_1
\eeq
under $\mc{TM}$, which means $(\mc{TM})^2=-1$ for the monopole.

After gauging, we identify the electric charge of the resulting $U(1)$ QSL as the fermions of the TCI, and the magnetic monopole as the monopole of the TCI. Then the charge is a Kramers singlet under $\mc{T}$, and it has $Z^2=-1$. The monopole has $\mc{M}^2=-1$ and $Z^2=-1$. So this $U(1)$ QSL is denoted as $(E_{fZ}M_{bMZ})_-$.

\subsection{Kramers doublet fermions} \label{app: sub-doublet}

Next consider the $\mc{T}$ action
\beq
\mc{T}: \psi\rightarrow i\sigma_y\psi
\eeq
Notice, combining this and (\ref{eq: MTCI-symmetries}), we see $Z^2\equiv(\mc{TM})^2=1$ for the fermions. This generalization of MTCI will be labeled as TMTCI-2.

Under $\mc{T}$ the monopoles transform as in (\ref{eq: WTI-T}), and under $\mc{TM}$
\beq
M_1\rightarrow-M_1,
\quad
M_2\rightarrow-M_2
\eeq
so $(\mc{TM})^2=1$ on the monopoles.

After gauging, we identify the electric charge of the resulting $U(1)$ QSL as the fermions of the TCI, and the magnetic monopole as the monopole of the TCI. Then the charge is a Kramers doublet under $\mc{T}$, and it has $Z^2=1$. The monopole has $\mc{M}^2=-1$ and $Z^2=1$. So this $U(1)$ QSL is denoted as $(E_{fT}M_{bM})_-$.

In this appendix we have derived the quantum numbers of the monopoles from a symmetric gapless surface state of these TCIs, and it is instructive to reproduce these results by studying a symmetric STO of them. This will be done in Appendix \ref{app: STO-double-FKM}.

\section{Surface topological orders of various TCIs} \label{app: STO-double-FKM}

In this appendix we construct symmetric STOs of the DTI insulator described in Sec. \ref{subsec: double-Fu-Kane-Mele}, and those of TMTCI-1 and TMTCI-2 discussed in Appendix \ref{app: generalized-mirror-TCI}. Our method is based on that in Ref. \cite{Isobe2015,Song2016,Cheng2017,Hong2017}.

\subsection{STO of the doubled topological insulator}

Let us begin with the STO of the DTI discussed in Sec. \ref{subsec: double-Fu-Kane-Mele}. Starting from the surface theory (\ref{eq: Dirac-cones}) together with the symmetry (\ref{eq: double-FKM-symmetries}), we can introduce an extra term to the Hamiltonian
\beq
\delta H=m(x)\psi^\dag\sigma_y\tau_y\psi
\eeq
where $m(x)=m_0{\rm sgn}(x)$ represents a mass domain wall. This term respects all symmetries, and it gaps out the surface except at $x=0$, which now hosts a pair of helical Dirac fermions. In fact, these helical Dirac fermions are identical to those in the edge state of a 2D topological insulator. Therefore, a DTI is also characterized by having a 2D topological insulator on its mirror plane. Recall that a WTI can be viewed as a stack of 2D topological insulators. This then implies that, starting from a WTI protected by time reversal and $T_z$, the translation symmetry along the $z$ direction, one can obtain a DTI by suitably breaking $T_z$ while preserving a mirror symmetry with respect to the $z=0$ plane, such that the $z=0$ plane hosts a 2D topological insulator.

\begin{figure}[h]
  \centering
  \includegraphics[width=0.36\textwidth]{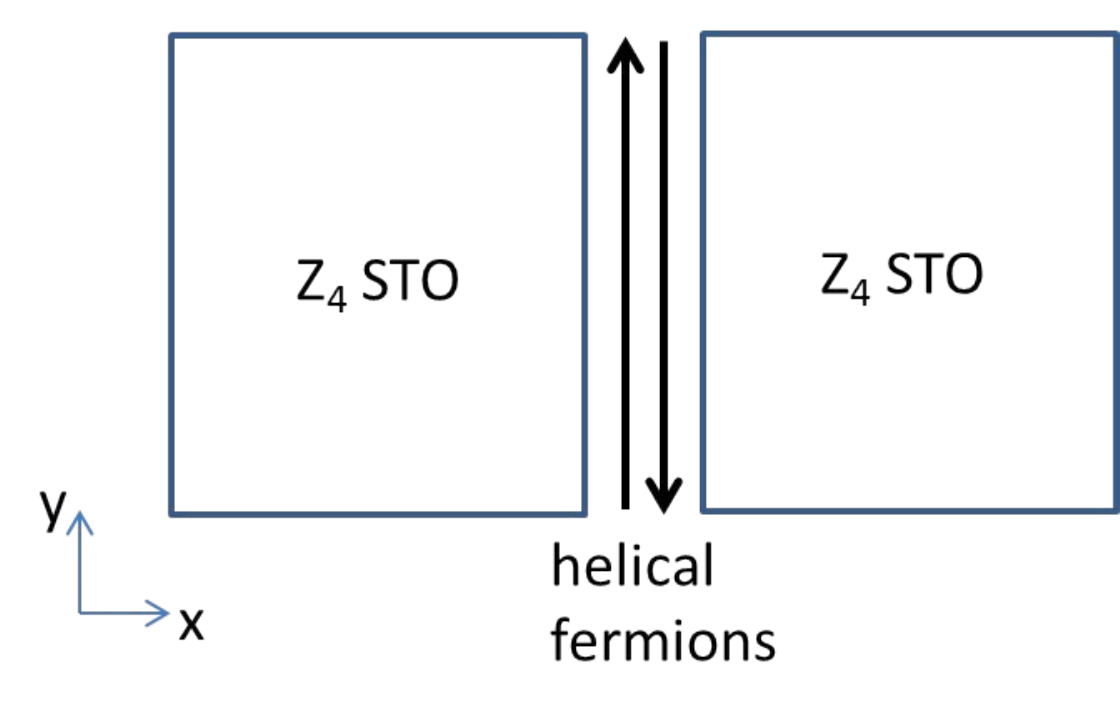}\\
  \caption{Gapping out the helical fermions by a $Z_4$ STO.}\label{fig: mirror-STO}
\end{figure}

In order to fully gap out the surface, we can introduce two mutually mirror symmetric $Z_4$ topological orders on the two sides of $x=0$, such that (the topological part of) the theory near $x=0$ is described by the following chiral Luttinger liquid theory\cite{Levin2012}
\beq \label{eq: chiral-Luttinger}
\mc{L}=\frac{K_{IJ}}{4\pi}\partial_t\Phi_I\partial_x\Phi_J
\eeq
where $\Phi=(\phi_{L1},\phi_{L2},\phi_{R1},\phi_{R2},\phi_{m1},\phi_{m2})^T$ satisfies the following Kac-Moody algebra
\beq
[\partial_y\Phi_I(y),\partial_{y'}\Phi_J(y')]=2\pi i(K^{-1})_{IJ}\partial_y\delta(y-y')
\eeq
In our case, the $K$-matrix is
\beq
K=
\left(
\begin{array}{ccc}
4\sigma_x&&\\
&4\sigma_x&\\
&&\sigma_z
\end{array}
\right)
\eeq
where the first (second) block represents the $Z_4$ topological order on the left (right) side of $x=0$, and the last block represents the helical fermion modes at $x=0$ (see Figure \ref{fig: mirror-STO}). Notice in defining the chiral Luttinger liquid theory, normal directions of the left and right STOs need to be chosen. In writing down the above $K$-matrix, we choose the normal direction for the left STO to be opposite to that for the right STO.

We assign the symmetries as
\beq \label{eq: Luttinger-symmetries}
\begin{split}
&(\phi_{L1},\phi_{L2},\phi_{R1},\phi_{R2},\phi_{m1},\phi_{m2})
\xrightarrow{U(1)}\\
&\
\left(\phi_{L1}-\frac{\theta}{2},\phi_{L2},\phi_{R1}-\frac{\theta}{2},\phi_{R2}, \phi_{m1}+\theta,\phi_{m2}+\theta\right),\\
&(\phi_{L1},\phi_{L2},\phi_{R1},\phi_{R2},\phi_{m1},\phi_{m2})
\xrightarrow{\mc{T}}\\
&\
\left(-\phi_{L1}+t\cdot\frac{\pi}{4},\phi_{L2},-\phi_{R1}+t\cdot\frac{\pi}{4},\phi_{R2}, -\phi_{m2},-\phi_{m1}+x\pi\right),\\
&(\phi_{L1},\phi_{L2},\phi_{R1},\phi_{R2},\phi_{m1},\phi_{m2})
\xrightarrow{\mc{M}}\\
&\quad\quad
\Big(\phi_{R1},\phi_{R2},\phi_{L1},\phi_{L2},\phi_{m1},\phi_{m2}+y\pi\Big)
\end{split}
\eeq
with $t=1$, $x=1$ and $y=0$. It is straightforward to check this is a consistent symmetry assignment. For example, one can check all local charge-1 objects are fermions, and they are Kramers doublets under $\mc{T}$ and also have $(\mc{TM})^2=-1$.

To gap out the helical fermions, consider introducing the following term to the Lagrangian (\ref{eq: chiral-Luttinger})
\beq \label{eq: Luttinger-gapping}
\begin{split}
\delta\mc{L}
=&-U\cos(4\phi_{L2}+4\phi_{R2}-2\phi_{m1}+2\phi_{m2})\\ &-V\cos(4\phi_{L1}+\phi_{m1}+\phi_{m2})\\
&-\eta V\cos(4\phi_{R1}+\phi_{m1}+\phi_{m2})
\end{split}
\eeq
with $\eta=1$, and $U$ and $V$ positive. It is straightforward to check that $\delta\mc{L}$ respects all symmetries, the arguments in the cosines in $\delta\mc{L}$ mutually commute, and pinning the values of these cosines by making $U$ and $V$ large does not break any symmetry spontaneously. Furthermore, in this strong-coupling limit the two $Z_4$ topological orders collapse into a single $Z_4$ topological order due to the coherent propagation of anyons across $x=0$. Therefore, this TCI can indeed have a symmetric $Z_4$ STO.

Now we take a closer look at the symmetry actions on the anyons of this $Z_4$ STO. We interpret $e^{i\phi_{L1}}$ and $e^{i\phi_{R1}}$ as the topological sector of the $Z_4$ charge $e$, and $e^{i\phi_{L2}}$ and $e^{i\phi_{R2}}$ as the topological sector of the $Z_4$ flux $m$. From the above symmetry assignment, the $e$ carries charge-1/2 under $U(1)$, while the $m$ is neutral. Based on the principle in Sec. \ref{subsec: monopole-STO}, $m^2$, which correspond to $e^{2i\phi_{L2}}$ and $e^{2i\phi_{R2}}$, are the surface avatars of the bulk monopole. To determine the quantum numbers of the monopole, we just need to determine the quantum numbers of $m^2$.

To check their quantum number under $\mc{M}$ and $\mc{TM}$, we use the method in Ref. \onlinecite{Qi2015}. First we consider a mirror symmetric string operator of this anyon
\beq \label{eq: string}
W=\exp\big(i\left(2\phi_{L2}+2\phi_{R2}-\phi_{m1}+\phi_{m2}\right)\big)
\eeq
The reason to consider this string operator is because it is able to coherently move an anyon $m^2$ from a point at $x<0$ to a point $x>0$, given that in the strong-coupling limit $\delta\mc{L}$ pins the value of $\la 4\phi_{L2}+4\phi_{R2}-2\phi_{m1}+2\phi_{m2}\ra=2\pi N$ with $N$ an integer. Under $\mc{M}$, this string operator does not change. This means $\mc{M}^2=1$ for this anyon. Under $\mc{TM}$,
\beq
W\rightarrow -\exp\big(-i\left(2\phi_{L2}+2\phi_{R2}-\phi_{m1}+\phi_{m2}\right)\big)
\eeq
In the strong-coupling limit, this becomes
\beq
W\rightarrow-W\la e^{-i(4\phi_{L2}+4\phi_{R2}-2\phi_{m1}+2\phi_{m2})}\ra =-W
\eeq
Therefore, $(\mc{TM})^2=-1$ for this anyon.

The above results are consistent with that obtained Sec. \ref{subsec: double-Fu-Kane-Mele} based on a symmetric gapless surface: the monopole has $\mc{M}^2=1$ and $(\mc{TM})^2=-1$. We note that other interesting properties of this STO can also be read off from the above construction, which is beyond the purpose of this paper and we refer interested readers to Ref. \onlinecite{Cheng2017,Hong2017}.

\subsection{STO of TMTCI-1}

Very similar as the above, a symmetric $Z_4$ STO can be constructed for TMTCI-1, and it is described by (\ref{eq: chiral-Luttinger}), (\ref{eq: Luttinger-symmetries}) and (\ref{eq: Luttinger-gapping}), but now with $t=0$, $x=0$, $y=1$ and $\eta=-1$. Again, the surface avatar of the bulk monopole corresponds to $e^{2i\phi_{L2}}$ and $e^{2i\phi_{R2}}$.

Now consider the string operator in (\ref{eq: string}). Under $\mc{M}$
\beq
W\rightarrow-W
\eeq
which means $\mc{M}^2=-1$ for the monopole. This is consistent with the result in Sec. \ref{subsec: mirror-TCI}. Under $\mc{TM}$
\beq
W\rightarrow -\exp\big(-i\left(2\phi_{L2}+2\phi_{R2}-\phi_{m1}+\phi_{m2}\right)\big)
\eeq
and in the strong-coupling limit we have $W\rightarrow-W$. Therefore, $(\mc{TM})^2=-1$ for the monopoles. This is consistent with the results obtained based on the symmetric gapless surface in Appendix \ref{app: sub-singlet}.

\subsection{STO of TMTCI-2}

TMTCI-2 can also have a symmetric $Z_4$ STO described by (\ref{eq: chiral-Luttinger}), (\ref{eq: Luttinger-symmetries}) and (\ref{eq: Luttinger-gapping}), but now with $t=1$, $x=1$, $y=1$ and $\eta=-1$. Again, the surface avatar of the bulk monopole corresponds to $e^{2i\phi_{L2}}$ and $e^{2i\phi_{R2}}$.

Now consider the string operator in (\ref{eq: string}). Under $\mc{TM}$ it becomes
\beq
W\rightarrow \exp\big(-i\left(2\phi_{L2}+2\phi_{R2}-\phi_{m1}+\phi_{m2}\right)\big)
\eeq
and in the strong-coupling limit we have $W\rightarrow W$. Therefore, $(\mc{TM})^2=1$ for the monopoles. This is consistent with the results obtained based on the symmetric gapless surface in Appendix \ref{app: sub-doublet}.

\bibliography{Monopole-TCI}

\end{document}